\input{aipcheck}

\documentclass[
    final            
  ]
  {aipproc}

\layoutstyle{6x9}

\begin{document}

\title{Particle Acceleration and Radiation associated
with Magnetic Field Generation from Relativistic
Collisionless Shocks}

\author{K.-I. Nishikawa}{
  address={National Space Science and Technology Center,
  Huntsville, AL 35805}
}

\author{P. Hardee}{
  address={Department of Physics and Astronomy,
  The University of Alabama, 
  Tuscaloosa, AL 35487}
}

\author{G. Richardson}{
  address={National Space Science and Technology Center,
  Huntsville, AL 35805}
}
\author{R. Preece}{
  address={Department of Physics, 
  University of Alabama in Huntsville,
  Huntsville, AL 35899 and National Space Science and Technology Center,
  Huntsville, AL 35805}}
\author{H. Sol}{
  address={LUTH, Observatore de Paris-Meudon, 5 place Jules Jansen 92195 
   Meudon Cedex, France}}
\author{G. J. Fishman}{
  address={NASA-Marshall Space Flight Center,
National Space Science and Technology Center,
  320 Sparkman Drive, SD 50, Huntsville, AL 35805}}

\begin{abstract}
Shock acceleration is an ubiquitous phenomenon in astrophysical plasmas. 
Plasma waves  and their associated instabilities (e.g., the Buneman 
instability, two-streaming instability, and the
Weibel instability) created in the shocks are responsible for particle 
(electron, positron, and ion) acceleration. Using a 3-D relativistic
electromagnetic particle (REMP) code, we have investigated particle
acceleration associated with a relativistic jet front propagating
through an ambient plasma with and without initial magnetic fields. We
find only small differences in the results between no ambient and weak
ambient magnetic fields. Simulations show that the Weibel instability
created in the collisionless shock front accelerates particles
perpendicular and parallel to the jet propagation direction.  The
simulation results show that this instability is responsible for
generating and amplifying highly nonuniform, small-scale magnetic fields, which
contribute to the electron's transverse deflection behind the jet
head.  The ``jitter'' radiation from deflected electrons
has different properties than synchrotron radiation which is calculated
in a uniform magnetic field. This jitter radiation may be
important to understanding the complex time evolution and/or spectral
structure in gamma-ray bursts, relativistic jets, and supernova
remnants.
 
\end{abstract}

\maketitle


\section{Introduction}

	The most widely known mechanism for the acceleration of particles 
in astrophysical environments usually with a power-law spectrum 
is Fermi acceleration. This mechanism for particle 
acceleration relies on the shock jump conditions at relativistic shocks 
(e.g., Gallant 2002). Most astrophysical shocks are collisionless since 
dissipation is dominated by wave-particle interactions rather than 
particle-particle collisions. Diffusive shock acceleration (DSA) relies on 
repeated scattering of charged particles by magnetic irregularities 
(Alfv\'en waves) to confine the particles near the shocks. However, particle 
acceleration near relativistic shocks is not due to DSA because the 
propagation of accelerated particles near shocks, in particular ahead of 
the shock, cannot be described as spatial diffusion. Anisotropies in the
angular distribution of the accelerated particles are large, and the 
diffusion approximation for spatial transport do not apply (Achterberg 
et al. 2001). Particle-in-cell (PIC) simulations may shed light on the physical 
mechanism of particle acceleration that involves the complicated dynamics 
of particles in relativistic shocks (Nishikawa et al. 2003, Silva et al. 
2003; Frederiksen et al. 2003a,b).

\section{Simulation model}

The simulations were performed using a $85 \times 85 \times 160$ grid with 
range of 55 to 85 million particles (27 particles$/$cell$/$species for 
the ambient plasma). Both periodic and radiating boundary conditions are 
used (Buneman 1993). The ambient electron and ion plasma has a mass ratio 
$m_{\rm i}/m_{\rm e} = 20$. The electron thermal velocity $v_{\rm e}$ is 
$0.1c$, where $c$ is the speed of light. The electron skin depth, 
$\lambda_{\rm ce} = c/\omega_{\rm pe}$, is $4.8\Delta$, where 
$\omega_{\rm pe} = (4\pi e^{2}n_{\rm e}/m_{\rm e})^{1/2}$ is the electron 
plasma frequency ($\Delta$ is the grid size). 

\section{Simulation results}

\subsection{Flat Magnetized jets}

The jet density of the flat jet is nearly $0.741n_{\rm b}$.  
The average jet velocity
$v_{\rm j} = 0.9798c$, and the Lorentz factor is 5 
(corresponds to 5 MeV). The time step
$\omega_{\rm pe}t = 0.026$, the ratio
$\omega_{\rm pe}/\Omega_{\rm e} = 2.89$, the Alfv\'en speed
$v_{\rm A} = 0.0775c$, and the Alfv\'en Mach number
$M_{\rm A} = v_{\rm j}/v_{\rm A} = 12.65$. The gyroradii of ambient
electrons and ions are $1.389\Delta$, and $6.211\Delta$, respectively.
In this case,
the jet makes contact with the ambient plasma at a 2D interface
spanning the computational domain. Therefore, the dynamics of the
jet head and the propagation of a shock in the downstream region are 
studied.  The Weibel instability is excited and the electron density is
perturbed as shown in Fig.\ 1a. 
\begin{figure}[h!]
\includegraphics{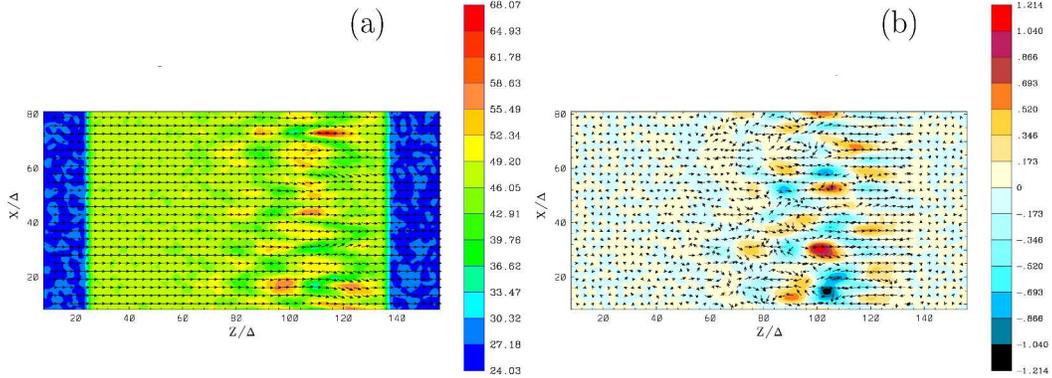}
\caption{\baselineskip 12pt The Weibel instability for the flat jet is 
illustrated in 2D images in the $x - z$ plane
($y = 43\Delta$) in the center of the jet.  In
(a) the colors indicate the electron density with magnetic fields represented
by arrows and in (b) the colors indicate the $y$-component of the current
density ($J_{\rm y}$) with $J_{\rm z}, J_{\rm x}$ indicated by the arrows.
The Weibel instability perturbs the electron density, leading to nonuniform
currents and highly structured magnetic fields.}
\end{figure}
The electrons are deflected by the perturbed (small) transverse
magnetic fields ($B_{\rm x}, B_{\rm y}$) via the Lorentz force:
$-e({\bf v} \times {\bf B})$, generating filamented current
perturbations ($J_{\rm z}$), which enhance the transverse magnetic
fields (Weibel 1959; Medvedev and Loeb 1999). The complicated current
structures due to the Weibel instability are shown in Fig.\ 1b. The
sizes of these structures are nearly the electron skin depth
($4.8\Delta$). This is in good agreement with $\lambda \approx
2^{1/4}c\gamma_{\rm th}^{1/2}/\omega_{\rm pe} \approx 1.188\lambda_{\rm
ce} = 5.7\Delta$ (Medvedev \& Loeb 1999). Here, $\gamma_{\rm th}$ is a
thermal Lorentz factor, and $\omega_{\rm pe}$ is the electron plasma
frequency. The shapes are elongated along the direction of the jet (the
$z$-direction, horizontal in Fig.\ 1).  

The growth rate of the Weibel instability is calculated to be, $\tau
\approx \gamma_{\rm sh}^{1/2}/\omega_{\rm pe} \approx 21.4$
($\gamma_{\rm sh} = 5$) (Medvedev \& Loeb 1999). This is in good
agreement with the simulation results with the jet head located at $z =
136\Delta$. Figure 1 suggests that the ``shock" has a thickness from
about $z/\Delta = 80 - 130$. Possibly, the ``turbulence" assumed for
diffusive shock acceleration corresponds to this shock region.
The width of the jet head is nearly the electron skin depth
($4.8\Delta$). The size of perturbations along the jet around $z =
120\Delta$ is nearly twice the electron skin depth. This result is
consistent with the previous simulations by Silva et al. (2003).  The
Weibel instability creates elongated shell-type structures which are
also shown in counter-streaming jet simulations (Nishikawa et al. 2003;
Silva et al. 2003;
Frederiksen et al. 2003a,b). The size of these structures transverse to
the jet propagation is nearly the electron skin depth ($4.8\Delta$).
Note that the size of the perturbations grows larger (see Fig.\ 1) behind
the jet front as smaller scale perturbations merge to larger sizes in
the nonlinear stage at the maximum amplitudes (Silva et al. 2003).

\section{Summary and Discussions}

     	 We have performed the first self-consistent, three-dimensional 
relativistic particle simulations of  electron-ion relativistic jets 
propagating through magnetized and unmagnetized electron-ion ambient 
plasmas. The Weibel instability is excited in the downstream region behind 
the jet head, where electron density perturbations and filamented currents 
are generated. The nonuniform electric field and magnetic field structures 
slightly decelerate the jet electrons and ions, while accelerating (heating) 
the jet electrons and ions in the transverse direction, in addition to accelerating 
the ambient material. The Weibel instability results from the 
fact that the electrons are deflected by the perturbed (small) transverse 
magnetic fields ($B_{\rm x}, B_{\rm y}$), and subsequently enhancement of the 
filamented current is seen  (Weibel 1959; Medvedev and Loeb 1999; Brainerd 2000; 
Gruzinov 2001). 

	The simulation results show that the initial jet kinetic energy goes 
to the magnetic fields and transverse acceleration of the jet particles through 
the Weibel instability. The properties of the synchrotron or ``jitter" 
emission from relativistic shocks are determined by the magnetic field 
strength, ${\bf B}$ and the electron energy distribution behind the shock. 
The following dimensionless parameters are used to estimate these values; 
$\epsilon_{\rm B} = U_{\rm B}/e_{\rm th}$ and $\epsilon_{\rm e} = 
U_{\rm e}/e_{\rm th}$ (Medvedev \& Loeb 1999). Here $U_{\rm B} = B^2/8\pi$, 
$U_{\rm e}$ are the magnetic and electron energy densities, and 
$e_{\rm th} = nm_{\rm i}c^{2}(\gamma_{\rm th} - 1)$ is 
the total thermal energy density behind the shock, where $m_{\rm i}$ is 
the ion mass, $n$ is the ion number density, and $\gamma_{\rm th}$ is the mean 
thermal Lorenz factor of ions. Based on the available diagnostics the following 
values are estimated; $\epsilon_{\rm B} \approx 0.02$ and $\epsilon_{\rm e}
\approx 0.3$. These estimates are made at the maximum amplitude ($z \approx
112\Delta$).

	Our present simulation study has provided the framework of the
fundamental dynamics of a relativistic shock generated within a relativistic 
jet. While some Fermi 
acceleration may occur at the jet front, the majority of electron 
acceleration takes place behind the jet front and cannot be characterized 
as Fermi acceleration.  Since the shock dynamics is complex and subtle, 
further comprehensive study is required for better understanding of the 
acceleration of electrons and the associated emission as compared with 
current theory (e.g., Rossi \& Rees 2002).  This further study will provide more 
insight into basic relativistic collisionless shock characteristics.  
The fundamental characteristics of such shocks are essential for a proper 
understanding of the prompt gamma-ray and afterglow emission in gamma-ray 
bursts, and also to an understanding of the particle reacceleration 
processes and emission from the shocked regions in relativistic AGN 
jets.

\begin{theacknowledgments}
K.N. is a NRC Senior Research Fellow at NASA Marshall Space Flight Center.
This research (K.N.) is partially supported by NSF ATM 9730230,
ATM-9870072, ATM-0100997, and INT-9981508. The simulations 
have been performed on ORIGIN 2000 and IBM p690 (Copper) at NCSA which 
is supported by NSF.
\end{theacknowledgments}


\bibliographystyle{aipproc}   



\end{document}